\documentclass[%
  preprint,
  amsmath,amssymb,
  aps,
]{revtex4-2}

\usepackage{graphicx}
\usepackage{dcolumn}
\usepackage{bm}
\usepackage{amsmath}
\usepackage{amssymb}
\usepackage{hyperref}
\usepackage{physics}  % for \bra, \ket, \mel
\DeclareMathOperator*{\argmin}{arg\,min}
\usepackage{xcolor}

\begin{document}

\preprint{Submitted to PRX Intelligence}

\title{Physics-Informed Classical and Quantum Neural Networks for
       One-Dimensional Schr\"{o}dinger Eigenvalue Problems}

\author{Tariq Mahmood}
\email{tariqmahmood.chep@pu.edu.pk}
\affiliation{Centre for High Energy Physics, University of the Punjab,
             Lahore, Pakistan}

\author{Waqas Arshad}
\email{waqasarshad9730@gmail.com}
\affiliation{Centre for High Energy Physics, University of the Punjab,
             Lahore, Pakistan}

\author{Bilal Naseer}
\email{Bilalnaseer1555@gmail.com}
\affiliation{Centre for High Energy Physics, University of the Punjab,
             Lahore, Pakistan}

\author{Alfredo Raya}
\email{alfredo.raya@umich.mx}
\affiliation{Facultad de Ingenier\'ia El\'ectrica, Universidad Michoacana de
             San Nicol\'as de Hidalgo, Morelia, Michoac\'an, M\'exico}
\affiliation{Centro de Ciencias Exactas, Universidad del B\'io-B\'io, Avda.
             Andr\'es Bello 720, Casilla 447, Chill\'an, Chile}

\date{\today}

\begin{abstract}
The Schr\"{o}dinger equation in one spatial dimension admits a small set of
exactly solvable potentials that serve as natural proving grounds for any new
eigenvalue solver. We formulate Physics-Informed Neural Networks (PINNs) and
Physics-Informed Quantum Neural Networks (PIQNNs) for the time-independent
Schr\"{o}dinger equation and apply them to three of these benchmarks: the
harmonic oscillator, the infinite square well, and the finite square well. In
each case a composite loss encodes the differential-equation residual, the
normalization condition, the boundary behavior, and the orthogonality between
eigenstates, so that the trial wave function is driven toward a genuine
eigenfunction without supervised data. The eigenvalues and wave functions
returned by both methods are compared against the exact spectra and against
three classical references: the matrix Numerov method, the finite difference
method, and the shooting method. For the smooth oscillator the two neural
solvers reproduce the lowest four eigenvalues to parts per million, while for
the square wells they recover the analytic levels with comparable fidelity even
where the potential is discontinuous. The quantum circuit, built as a layered
angle-embedding ansatz with strongly entangling blocks, converges more reliably
than its classical counterpart on the higher excited states, where the loss
landscape of the classical network becomes harder to navigate.
\end{abstract}

\keywords{Artificial Neural Network; Quantum Neural Network;
          Physics-Informed Neural Network; Physics-Informed Quantum Neural
          Network; Schr\"{o}dinger equation; quantum harmonic oscillator;
          square-well potential}

\maketitle

% ==================================================================
\section{\label{sec:intro}Introduction}
% ==================================================================

\begin{figure}[htbp]
  \centering
 \includegraphics[width=0.5\columnwidth]{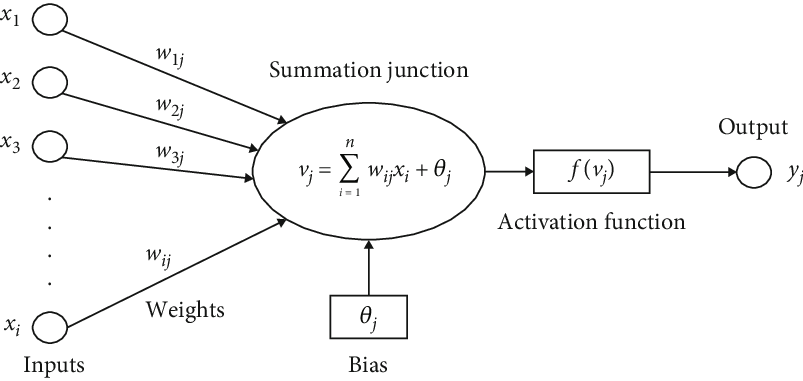}
  \caption{\label{fig:neuron}Working diagram of an artificial neuron.}
\end{figure}

\begin{figure}[htbp]
  \centering
\includegraphics[width=0.5\columnwidth]{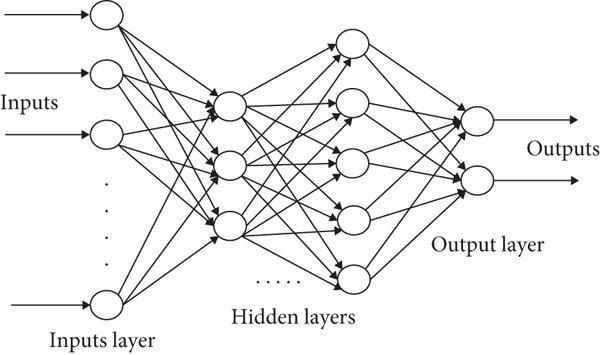}
  \caption{\label{fig:ann}Working diagram of a multilayer artificial
           neural network.}
\end{figure}

The architecture of artificial neural networks (ANNs) traces its conceptual
origin to the model of interconnected binary units proposed by McCulloch and
Pitts in 1943~\cite{McCulloch1943}, in which the complex biochemistry of
neuronal signaling is distilled into two discrete states: active and resting.
From that minimal starting point, the field has expanded into a broad and
heterogeneous landscape of models~\cite{Basheer2000,Qamar2023}. The basic
operational logic of an artificial neuron and a multilayer network is
illustrated schematically in Figs.~\ref{fig:neuron} and \ref{fig:ann},
respectively. One of the most consequential early applications of these ideas
to physics was the direct use of neural networks to solve differential
equations. Lee and Kang~\cite{Lee1990} demonstrated in 1990 that a feedforward
network can be trained to satisfy a differential operator together with its
boundary and initial conditions, thereby converting the problem of finding a
solution into a parameter optimization problem. Lagaris, Likas, and
Fotiadis~\cite{Lagaris1998} developed this idea further in 1998 by embedding the
boundary conditions directly into the trial solution, so that the network is
freed from enforcing them explicitly and can concentrate entirely on satisfying
the governing equation. The resulting approach, now recognized as the prototype
of physics-informed neural solvers, was shown to generalize better than
finite-element methods on several benchmark problems. The modern formulation,
in which the residual of the governing equation is minimized at a set of
collocation points, was placed on a general footing by Raissi, Perdikaris, and
Karniadakis~\cite{Raissi2019} and has since grown into the wide methodology
reviewed by Karniadakis and co-workers~\cite{Karniadakis2021}. Closely related
deep-learning treatments of differential equations include the Galerkin-type
scheme of Sirignano and Spiliopoulos~\cite{Sirignano2018} and the
high-dimensional solver of Han, Jentzen, and E~\cite{HanJentzenE2018}.

\begin{figure}[htbp]
  \centering
\includegraphics[width=0.5\columnwidth]{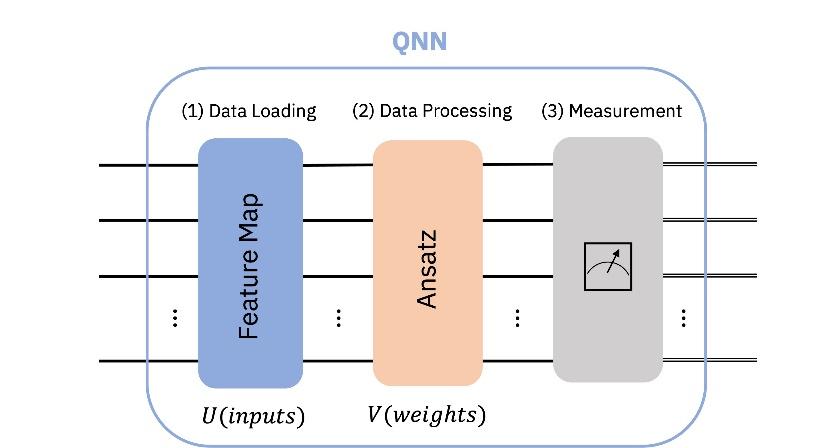}
  \caption{\label{fig:qnn}Schematic of a parameterized quantum neural
           network circuit~\cite{Qiskit}.}
\end{figure}

Quantum machine learning brings an additional layer of computational potential
to this framework by exploiting two features of quantum mechanics that have no
classical analog: superposition, which allows a qubit to inhabit a continuous
family of states simultaneously, and entanglement, which creates non-local
correlations between qubits that are impossible to reproduce efficiently on
classical hardware. These properties underpin the computational advantages that
have been identified and explored in the interdisciplinary
literature~\cite{DasSarma2019,Dunjko2017,Biamonte2017,Cerezo2022,Dawid2022,
InfoTheoretic2021}. A quantum neural network (QNN) is in essence a parameterized
quantum circuit organized into three segments: a data-encoding subcircuit that
maps classical input into a quantum state, a variational subcircuit whose
parameters are optimized during training, and a measurement layer whose output
is fed back to a classical optimizer~\cite{Wang2021}. The elementary gates
available in this framework include single-qubit rotations and multi-qubit
entangling operations such as the Hadamard, Pauli, CNOT, and Toffoli gates. The
concept of a quantum computer was introduced by Feynman~\cite{Feynman2018} and
the first concrete quantum neural network model was proposed by
Kak~\cite{Kak1995} in 1995. Andrecut and Ali~\cite{Andrecut2002} subsequently
showed in 2002 how learning and classification tasks can be formulated entirely
in terms of unitary quantum gates. A representative circuit architecture is
shown schematically in Fig.~\ref{fig:qnn}.

In high-energy physics (HEP), where experiments at the Large Hadron Collider and
facilities around the world probe the fundamental interactions described by the
Standard Model~\cite{CERN2023a,Radovic2018,CERN2023b}, machine learning has
become an indispensable tool. Neural networks were applied early on to particle
identification through calorimeter topology~\cite{Damazio2004}, and their use
has since expanded to cover classification~\cite{Therhaag2013,Teodorescu2008},
searches for the Higgs boson~\cite{Sadowski2014,Baldi2016}, studies of
hadron-hadron collisions~\cite{Radi2013}, and the spectroscopy of exotic
states~\cite{Mutuk2019a,Mutuk2019b,Mutuk2019c}. In parallel, neural methods have
been brought to bear on the Schr\"{o}dinger equation itself.
Miranker~\cite{Miranker2006} derived a neural-network wave formalism in 2006
that recovers the time-dependent Schr\"{o}dinger equation as a limiting case.
Hermann, Sch\"{a}tzle, and No\'e~\cite{Hermann2020} solved the many-electron
Schr\"{o}dinger equation with a deep network capable of representing the
ground-state energy of atoms from first principles, an approach paralleled by
the FermiNet ansatz of Pfau and collaborators~\cite{Pfau2020}. Mutuk has applied
neural methods to one-dimensional anharmonic oscillators~\cite{Mutuk2018a}, to
Cornell potentials~\cite{Mutuk2019a}, and to heavy quarkonia mass
spectra~\cite{Mutuk2018b}. Lema and Arizmendi~\cite{Lema2020} demonstrated that
an unsupervised network can recover ground-state energies and wave functions for
a particle in a box. More recently, Mahmood \textit{et al.}~\cite{Mahmood2024}
solved the radial Schr\"{o}dinger equation for charmonium with an ANN,
outperforming standard numerical methods in generality and continuity. The case
that sits closest to the present study is that of Jin, Mattheakis, and
Protopapas~\cite{Jin2022}, who used an unsupervised network with an
orthogonality penalty to recover the spectra of the infinite well and the
harmonic oscillator, and the parametric extension of that idea to molecular
systems by Mattheakis \textit{et al.}~\cite{Mattheakis2022}.

The emergence of QNNs has sharpened the case for going beyond classical
networks. Because a parameterized quantum circuit operates over the full Hilbert
space of its qubits, it can represent correlations that scale exponentially with
system size, a regime where classical networks become prohibitively expensive.
Schuld, Sinayskiy, and Petruccione~\cite{Schuld2015} provided an early
systematic overview of these possibilities in 2015, and Havl\'i\v{c}ek
\textit{et al.}~\cite{Havlicek2019} confirmed the utility of quantum feature maps
in supervised learning settings. Biamonte \textit{et al.}~\cite{Biamonte2017}
gave a comprehensive treatment of where quantum machine learning can offer
concrete advantages over classical approaches, and the broader family of
variational quantum algorithms within which a PIQNN sits is surveyed by Cerezo
\textit{et al.}~\cite{Cerezo2021}. Together, these studies motivate extending the
physics-informed methodology to quantum circuits.

This paper presents a unified PINN and PIQNN framework for the time-independent
Schr\"{o}dinger equation, validated on three of the textbook potentials whose
spectra are known analytically or follow from a simple transcendental
condition: the harmonic oscillator, the infinite square well, and the finite
square well. The first is smooth and rewards a solver that captures gentle
curvature, while the latter two are piecewise constant and force the wave
function to develop sharp features at the well edges, so that together they
exercise rather different aspects of an approximator. The framework is described
in detail in Sec.~\ref{sec:method}; the harmonic oscillator results appear in
Sec.~\ref{sec:results}; the square-well results appear in Sec.~\ref{sec:wells};
and Sec.~\ref{sec:conclusion} discusses the outlook for applying the same
methodology to realistic bound-state systems.

% ==================================================================
\section{\label{sec:method}Methodology}
% ==================================================================

% ------------------------------------------------------------------
\subsection{Classical numerical benchmarks}
% ------------------------------------------------------------------

Three established numerical methods serve as benchmarks throughout this study.
The matrix Numerov method~\cite{Pillai2012} reformulates the Schr\"{o}dinger
equation as a generalized matrix eigenvalue problem using a fourth-order
finite-difference scheme; its accuracy scales with the number of grid points and
is well understood for a wide range of potentials. The finite-difference method
developed by Simos~\cite{Simos1997,Simos1998} constructs discrete approximations
with controlled phase-lag properties that improve accuracy for oscillatory
solutions. The shooting method~\cite{Lambert2001,Gulyamov2022} converts the
boundary-value problem into an initial-value problem and iteratively adjusts an
energy guess until the wave function satisfies the boundary condition at both
ends of the domain. Each of these methods produces reliable results when the
spatial grid is sufficiently fine, but all three require manual tuning of
discretization parameters and scale poorly with the number of spatial dimensions
or the complexity of the potential.

% ------------------------------------------------------------------
\subsection{Physics-Informed Neural Network (PINN)}
% ------------------------------------------------------------------

The PINN approach represents the eigenfunction $\psi(x)$ through an ansatz that
encodes the correct asymptotic behavior analytically, leaving the network
responsible only for the interior structure of the solution. The trial wave
function is written as
\begin{equation}\label{eq:ansatz_pinn}
  u_{\text{trial}}(x) = x \, B(x) \, N_{\text{ANN}}(x),
\end{equation}
where the prefactor $x$ enforces continuity and differentiability at the origin,
$B(x) = e^{-\gamma x^2}$ is an envelope that guarantees exponential decay in the
asymptotic region, and $N_{\text{ANN}}(x)$ is the output of a feedforward neural
network with trainable parameters $\theta$. The Hamiltonian operator acts on the
trial wave function as
\begin{equation}
  \hat{H}\,u_{\text{trial}}(x) = -\frac{\hbar^2}{2m}
    \frac{d^2 u_{\text{trial}}}{dx^2} + V(x)\,u_{\text{trial}}(x),
\end{equation}
and the physical residual at a collocation point $x_i$ is
\begin{equation}\label{eq:residual}
  R(x_i;\,\theta,\epsilon) = \hat{H}\,u_{\text{trial}}(x_i)
    - \epsilon\,u_{\text{trial}}(x_i),
\end{equation}
where $\epsilon$ is either a trainable scalar or is evaluated at each step via
the Rayleigh quotient~\cite{Griffiths2018}
\begin{equation}\label{eq:rayleigh}
  \epsilon = \frac{\displaystyle\int \psi^*(x)\,\hat{H}\,\psi(x)\,dx}
                   {\displaystyle\int |\psi(x)|^2\,dx}.
\end{equation}

The network is trained by minimizing a composite loss function that
simultaneously enforces four physical conditions. The PDE residual loss,
\begin{equation}\label{eq:lpde}
  \mathcal{L}_{\text{PDE}} = \frac{1}{N_x}
    \sum_{i=1}^{N_x} R(x_i;\,\theta,\epsilon)^2,
\end{equation}
which measures how closely the trial function satisfies the eigenvalue equation at
$N_x$ collocation points in the interior of the domain. The normalization loss,
\begin{equation}\label{eq:lnorm}
  \mathcal{L}_{\text{norm}} =
    \left(\int |u_{\text{trial}}(x)|^2\,dx - 1\right)^2,
\end{equation}
that ensures  the correct normalization of the  wave function. The boundary loss,
\begin{equation}\label{eq:lbc}
  \mathcal{L}_{\text{BC}} = \frac{1}{N_b}
    \sum_{j=1}^{N_b} \left[u_{\text{trial}}\!\left(x_j^{(b)}\right)\right]^2,
\end{equation}
which penalizes non-zero values at the $N_b$ boundary points, typically at the edges
of the computational domain where the wave function must vanish. For excited
states, the orthogonality loss,
\begin{equation}\label{eq:lortho}
  \mathcal{L}_{\text{ortho}} = \sum_{k=0}^{n-1}
    \left(\int u_{\text{trial}}(x)\,u_k(x)\,dx\right)^2,
\end{equation}
penalizes overlap with the already-converged lower eigenstates $u_k$, which
prevents the optimizer from collapsing to a previously found state. The total
training objective is the weighted sum
\begin{equation}\label{eq:ltotal}
  \mathcal{L}_{\text{total}} = \lambda_{\text{PDE}}\,\mathcal{L}_{\text{PDE}}
    + \lambda_{\text{norm}}\,\mathcal{L}_{\text{norm}}
    + \lambda_{\text{BC}}\,\mathcal{L}_{\text{BC}}
    + \lambda_{\text{ortho}}\,\mathcal{L}_{\text{ortho}},
\end{equation}
and the optimal network parameters and eigenvalue are found by
\begin{equation}\label{eq:optim}
  (\theta^*,\epsilon^*) = \argmin_{\theta,\epsilon}\,
    \mathcal{L}_{\text{total}}(\theta,\epsilon).
\end{equation}
Training is carried out with the Adam optimizer; the weights
$\lambda_{\text{PDE}}=1.0,\, \lambda_{\text{norm}} =5.0,
\, \lambda_{\text{BC}}=\lambda_{\text{ortho}}=10.0$ are treated as
hyperparameters. The network architecture comprises a single hidden layer with
$\tanh$ activation function, 32--64 units and a learning rate $l_r=0.001$, with
5000 epochs. The collocation points range from $-10$ to $10$ with 800 grid
points. Upon convergence, the network simultaneously provides an approximation
to the eigenfunction and an estimate of the corresponding eigenvalue.
%%%%%%%%%%%%%%%%%%%%%%%%%%%%%%%%%%%%%%%%%%%%%%%%%%%%%%%%%%%%
% NOTE TO AUTHORS: the architecture above (single hidden layer, 32-64 units,
% tanh, lr=1e-3, 5000 epochs, 800 collocation points, lambda = 1/5/10/10) was
% carried over from the working notes. Please confirm these match the final
% production runs, and add N_b (number of boundary points) explicitly, before
% submission. PRX Intelligence requires a complete reproducibility statement.
%%%%%%%%%%%%%%%%%%%%%%%%%%%%%%%%%%%%%%%%%%%%%%%%%%%%%%%
% ------------------------------------------------------------------
\subsection{Physics-Informed Quantum Neural Network (PIQNN)}
% ------------------------------------------------------------------

\begin{figure}[htbp]
  \centering
\includegraphics[width=0.5\columnwidth]{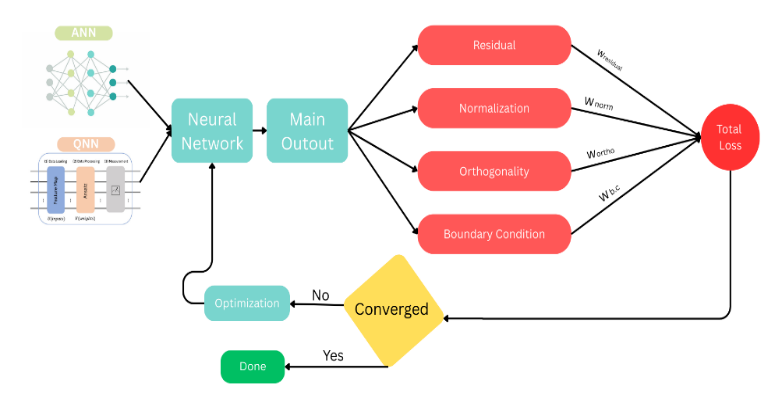}
  \caption{\label{fig:flowchart}Flow diagram illustrating the PIQNN
  architecture and training workflow. The circuit encodes the spatial coordinate
  into a quantum state, applies a sequence of parameterized entangling layers,
  and extracts the trial wave function via an expectation value. The resulting
  output feeds into the composite physics-informed loss, which drives both the
  circuit parameters and the eigenvalue estimate toward convergence.}
\end{figure}

The PIQNN formulation replaces the classical feedforward network in
Eq.~(\ref{eq:ansatz_pinn}) with a parameterized quantum circuit, a construction
that belongs to the family of variational quantum
algorithms~\cite{Cerezo2021,Mitarai2018}. The ansatz retains the same structure,
\begin{equation}\label{eq:ansatz_piqnn}
  U(x) = x\,e^{-\gamma x^2}\,N_{\text{QNN}}(x),
\end{equation}
where $N_{\text{QNN}}(x)$ is the output of the quantum circuit with trainable
parameters $\theta$. The overall architecture and training workflow are depicted
in Fig.~\ref{fig:flowchart}.

The classical coordinate $x$ is mapped to a quantum register of $n_q$ qubits
through angle embedding,
\begin{equation}\label{eq:embed}
  \phi(x) = \bigotimes_{i=1}^{n_q} R_y(x)\ket{0},
\end{equation}
where $R_y(\alpha) = e^{-i\alpha\sigma_y/2}$ is a rotation gate around the
$y$-axis. The variational part of the circuit consists of $L$ entangling layers,
\begin{equation}\label{eq:circuit}
  U(\theta) = \prod_{l=1}^{L} \mathcal{E}_l(\theta_l),
\end{equation}
where $\mathcal{E}_l(\theta_l)$ is the $l$-th entangling block, comprising
single-qubit rotations and CNOT gates arranged to generate correlations across
all qubits. The QNN output is the expectation value of the tensor product of
Pauli-$Z$ operators,
\begin{equation}\label{eq:qnn_out}
  N_{\text{QNN}}(x) = \bra{0}U^\dagger(\theta)\,
    \hat{O}\,U(\theta)\ket{0}, \quad
    \hat{O} = \bigotimes_{i=1}^{n_q} \sigma_z^{(i)}.
\end{equation}
This construction maps a real classical input to a real output while maintaining
full differentiability with respect to $\theta$ through the parameter-shift
rule~\cite{SchuldGrad2019}, which enables gradient-based optimization compatible
with the Adam optimizer used for the classical PINN.

The training objective for the PIQNN is structurally identical to
Eq.~(\ref{eq:ltotal}),
\begin{equation}\label{eq:loss_qnn}
  \mathcal{L} = \bigl\langle\bigl(\hat{H}\,U(x) -
    \epsilon\,U(x)\bigr)^2\bigr\rangle
    + \lambda_1\!\left(\int|U(x)|^2dx - 1\right)^2
    + \lambda_2\,U(x_{\max})^2
    + \lambda_3\!\sum_k\!\left(\int U(x)\,u_k(x)\,dx\right)^2,
\end{equation}
where the four terms enforce the Schr\"{o}dinger equation residual,
normalization, decay at the boundary $x_{\max}$, and orthogonality to previously
converged eigenstates, respectively. The quantum circuit is built from four
qubits arranged in three strongly entangling layers, optimized with Adam, and
implemented in PyTorch coupled to PennyLane~\cite{Bergholm2018}.
%%%%%%%%%%%%%%%%%%%%%%%%%%%%%%%%%%%%%%%%%%%%%%%%
% NOTE TO AUTHORS: please confirm the entanglement topology (PennyLane
% StronglyEntanglingLayers is all-to-all per layer via cyclic CNOTs), the
% learning rate for the quantum parameters, and the number of measurement
% shots (or whether the analytic/state-vector backend was used), so the
% reproducibility statement is complete.
%%%%%%%%%%%%%%%%%%%%%%%%%%%%%%%%%%%%%%%%

% ==================================================================
\section{\label{sec:results}The Quantum Harmonic Oscillator}
% ==================================================================

% ------------------------------------------------------------------
\subsection{Problem setup}
% ------------------------------------------------------------------

The one-dimensional quantum harmonic oscillator,
\begin{equation}\label{eq:sho}
  \left[-\frac{\hbar^2}{2m}\frac{d^2}{dx^2}
    + \frac{1}{2}m\omega^2 x^2\right]\psi(x) = E\,\psi(x),
\end{equation}
provides an ideal validation case because its spectrum and wave functions are
known in closed form. In atomic units ($\hbar = m = \omega = 1$), the exact
eigenvalues are
\begin{equation}\label{eq:exact_evals}
  E_n = n + \tfrac{1}{2}, \qquad n = 0, 1, 2, \ldots,
\end{equation}
and the corresponding normalized eigenfunctions are
\begin{equation}\label{eq:exact_wfs}
  \psi_n(x) = \frac{1}{\sqrt{2^n n!}}\,\pi^{-1/4}\,H_n(x)\,e^{-x^2/2},
\end{equation}
where $H_n(x)$ denotes the physicist's Hermite polynomials.

The computational domain spans $x \in [-10, 10]$ and is discretized into
$N = 800$ evenly spaced grid points,
\begin{equation}
  x_i = x_{\min} + i\,\Delta x, \quad
  \Delta x = \frac{2\,|x_{\max}|}{N-1}, \quad
  i = 0, 1, \ldots, N-1.
\end{equation}
This resolution is more than sufficient to resolve the nodal structure of the
first four eigenstates. Both the PINN and PIQNN treat the eigenvalue $\epsilon$
as a trainable parameter optimized jointly with the network or circuit
parameters.

% ------------------------------------------------------------------
\subsection{Results}
% ------------------------------------------------------------------

Table~\ref{tab:evals} collects the eigenvalues predicted by each method for
quantum numbers $n = 0, 1, 2, 3$. All approaches recover the exact values to
within a few parts in $10^4$ or better, confirming that the problem is well-posed
and that each method has converged. The absolute errors in the PINN eigenvalues
range from $9 \times 10^{-6}$ to $1 \times 10^{-5}$, while those of the PIQNN
range from $1 \times 10^{-5}$ to $5.8 \times 10^{-4}$. The classical methods fall
within a comparable window, with the shooting method accurate to roughly
$10^{-5}$ and the finite difference method accumulating a slightly larger error
of order $10^{-4}$ for $n = 3$.

\begin{table}[htbp]
\caption{\label{tab:evals}Low-lying energy eigenvalues of the one-dimensional
quantum harmonic oscillator in atomic units. Exact values follow from
Eq.~(\ref{eq:exact_evals}).}
\begin{ruledtabular}
\begin{tabular}{ldddd}
  Method & \multicolumn{1}{c}{$n=0$} & \multicolumn{1}{c}{$n=1$}
         & \multicolumn{1}{c}{$n=2$} & \multicolumn{1}{c}{$n=3$} \\
  \hline
  Exact                       & 0.500000 & 1.500000 & 2.500000 & 3.500000 \\
  PINN                        & 0.500009 & 1.500010 & 2.500010 & 3.500008 \\
  PIQNN                       & 0.500580 & 1.500020 & 2.500070 & 3.500010 \\
  Shooting method             & 0.499990 & 1.499990 & 2.499960 & 3.499940 \\
  Matrix Numerov method       & 0.500020 & 1.500020 & 2.500020 & 3.500020 \\
  Finite difference method    & 0.499980 & 1.499902 & 2.499745 & 3.499510 \\
\end{tabular}
\end{ruledtabular}
\end{table}

\begin{figure}[htbp]
  \centering
  \includegraphics[width=0.5\columnwidth]{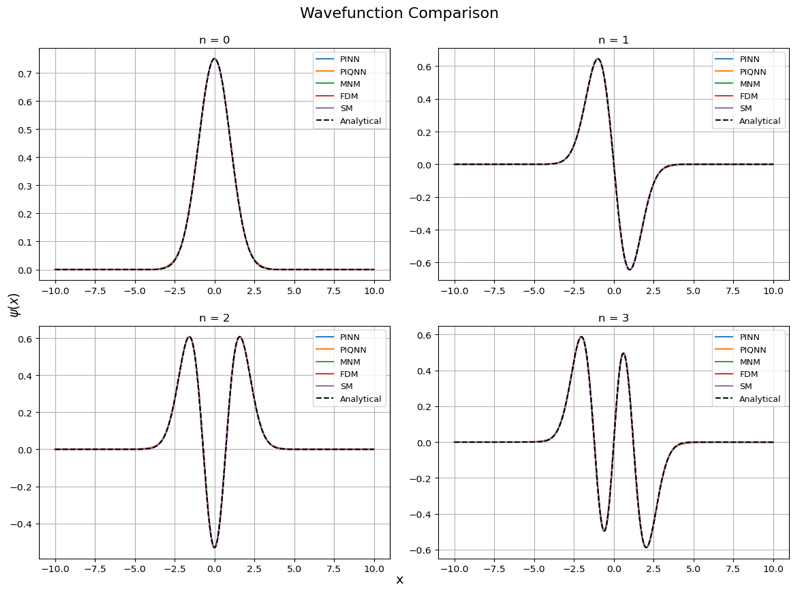}
  \caption{\label{fig:compare_all}Normalized wave functions of the
  one-dimensional quantum harmonic oscillator for quantum numbers
  $n = 0, 1, 2, 3$, computed by the exact analytical formula, the matrix Numerov
  method, the finite difference method, the shooting method, the PINN, and the
  PIQNN. All six curves are visually indistinguishable on this scale, confirming
  the agreement reported in Table~\ref{tab:evals}.}
\end{figure}

\begin{figure}[htbp]
  \centering
\includegraphics[width=0.5\columnwidth]{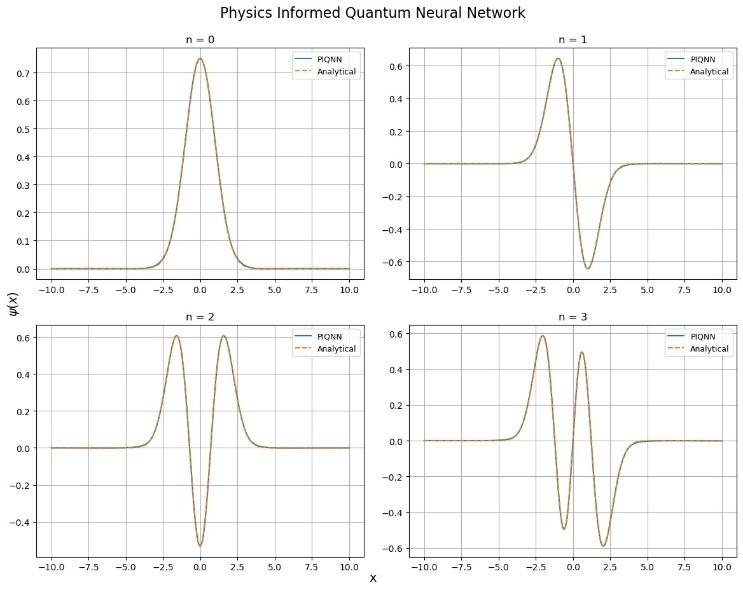}
  \caption{\label{fig:pinn_wf}Comparison of the PINN wave functions with the
  exact analytical eigenfunctions for $n = 0, 1, 2, 3$.}
\end{figure}

\begin{figure}[htbp]
  \centering
\includegraphics[width=0.5\columnwidth]{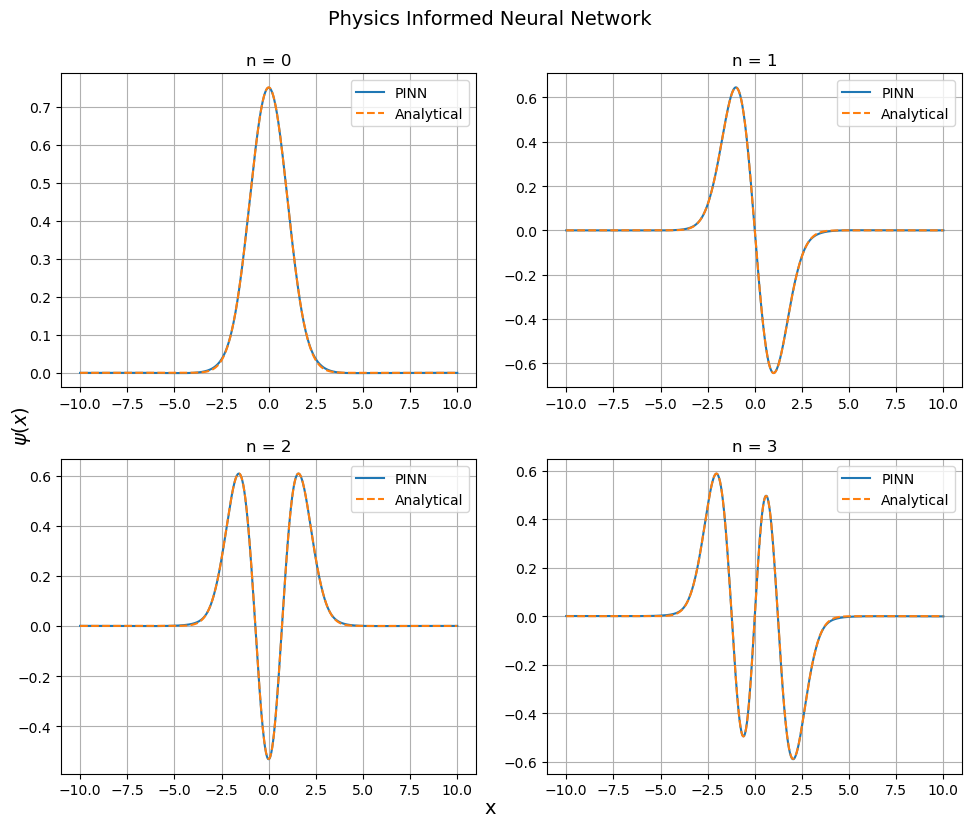}
  \caption{\label{fig:piqnn_wf}Comparison of the PIQNN wave functions with the
  exact analytical eigenfunctions for $n = 0, 1, 2, 3$.}
\end{figure}

The wave functions for $n = 0$ through $n = 3$ are shown in
Figs.~\ref{fig:compare_all}--\ref{fig:piqnn_wf}. Figure~\ref{fig:compare_all}
places all six methods on the same axes, making it straightforward to assess the
global agreement. Figures~\ref{fig:pinn_wf} and \ref{fig:piqnn_wf} isolate the
PINN and PIQNN predictions respectively, allowing a closer inspection of the
phase and amplitude of each eigenstate.

% ------------------------------------------------------------------
\subsection{Discussion}
% ------------------------------------------------------------------

The classical numerical methods recover accurate eigenvalues when the spatial
grid is fine enough, but their performance is sensitive to the choice of step
size and domain boundary, and errors grow noticeably for $n = 2$ and $n = 3$ in
the finite difference approach (Table~\ref{tab:evals}). By contrast, both neural
methods operate on a mesh-free domain: the physics is embedded through the
composite loss rather than through a discretization scheme. This means that
adding more collocation points is computationally cheaper than refining a spatial
grid in multiple dimensions, and that the method generalizes naturally to domains
where structured grids are awkward.

The more consequential difference between the two neural approaches concerns
their behavior on excited states. The PINN, like any classical neural network,
has a limited number of independent parameters and can be drawn into local minima
of the loss landscape, particularly when the orthogonality penalty for higher
states competes with the PDE residual term. The PIQNN partially circumvents this
difficulty through the geometry of its parameter space: because the quantum
circuit explores a continuous manifold of unitary transformations over the full
Hilbert space of its qubits, it retains expressivity even as the number of nodes
in the target wave function increases. The improved convergence of the PIQNN for
$n = 2$ and $n = 3$ reflects this difference in practice, although a quantitative
characterization of the landscape, including the onset of barren
plateaus~\cite{McClean2018} as the circuit grows, would be needed to make the
statement rigorous.

% ==================================================================
\section{\label{sec:wells}Square-Well Potentials}
% ==================================================================

The harmonic oscillator is a smooth potential whose eigenfunctions decay as
Gaussians, so it rewards an approximator that captures gentle curvature. A
complementary and in some respects harder test is provided by potentials that
change abruptly in space, where the wave function must develop sharp features at
the well edges. The infinite and finite square wells supply exactly this kind of
test, and both belong to the standard repertoire of solvable one-dimensional
problems treated in introductory quantum mechanics~\cite{Griffiths2018}. The
same systems were used by Jin, Mattheakis, and Protopapas~\cite{Jin2022} to
validate an unsupervised neural eigensolver, which makes them a useful point of
contact with the wider physics-informed literature.

% ------------------------------------------------------------------
\subsection{Infinite square well}
% ------------------------------------------------------------------

A particle of mass $m$ confined to the interval $[0, L]$ by an infinitely deep
well satisfies the free-particle equation inside the box together with the
hard-wall conditions $\psi(0) = \psi(L) = 0$. The normalized eigenfunctions and
eigenvalues are
\begin{equation}\label{eq:infwell}
  \psi_n(x) = \sqrt{\frac{2}{L}}\,\sin\!\left(\frac{n\pi x}{L}\right),
  \qquad
  E_n = \frac{n^2 \pi^2 \hbar^2}{2 m L^2}, \quad n = 1, 2, 3, \ldots,
\end{equation}
so that the spectrum grows quadratically with $n$ and the levels are
non-degenerate. In atomic units with $L = 1$ they reduce to
$E_n = n^2 \pi^2 / 2$. Table~\ref{tab:infwell} lists the first four levels
together with the values returned by the two neural solvers and the three
classical references. With the exception of one entry flagged in the note below,
every method reproduces the analytic levels to within a few parts in $10^4$,
which shows that the abrupt confinement does not, by itself, degrade the
physics-informed solution.

\begin{table}[htbp]
\caption{\label{tab:infwell}Lowest four energy levels of the one-dimensional
infinite square well in atomic units ($\hbar = m = L = 1$). Exact values
follow from Eq.~(\ref{eq:infwell}).}
\begin{ruledtabular}
\begin{tabular}{ldddd}
  Method & \multicolumn{1}{c}{$n=1$} & \multicolumn{1}{c}{$n=2$}
         & \multicolumn{1}{c}{$n=3$} & \multicolumn{1}{c}{$n=4$} \\
  \hline
  Exact                       & 4.934802 & 19.739209 & 44.413220 & 78.956835 \\
  PINN                        & 4.934899 & 19.739101 & 44.412971 & 78.956360 \\
  PIQNN                       & 4.934898 & 19.739132 & 44.413147 & 78.460274\\
  Matrix Numerov method       & 4.934802 & 19.739209 & 44.413220 & 78.956835 \\
  Finite difference method    & 4.934796 & 19.739107 & 44.412705 & 78.955208 \\
  Shooting method             & 4.934802 & 19.739209 & 44.413220 & 78.956835 \\
\end{tabular}
\end{ruledtabular}
\end{table}
A comparison of the wavefunctions for the states with $n=0,1,2,3$ with the different techniques is illustrated
in Fig.~\ref{fig:all_infinite}
\begin{figure}[htbp]
  \centering
\includegraphics[width=0.5\textwidth]{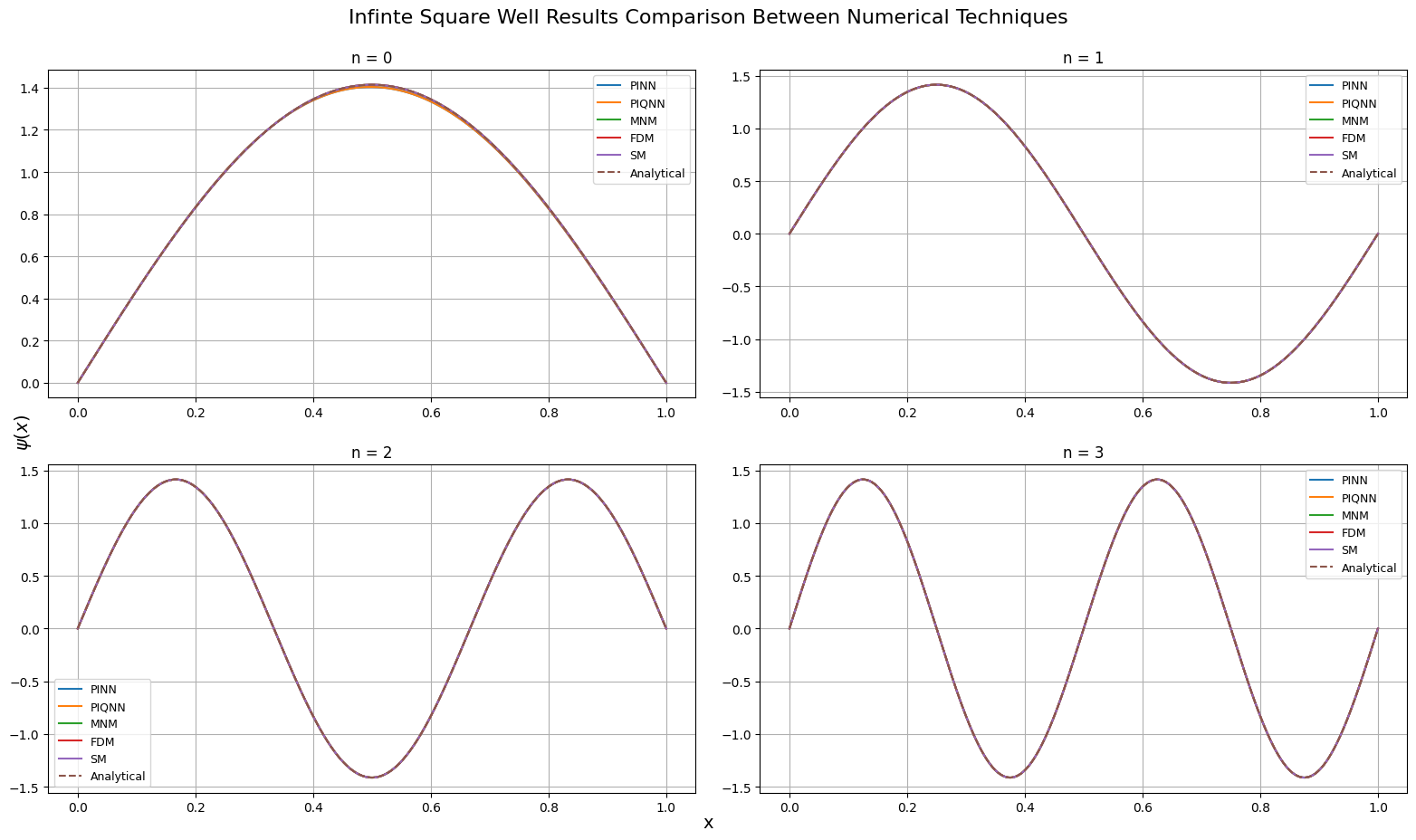}
  \caption{\label{fig:all_infinite}Comparison of the wave functions for the infinite square well with the different classical and quantum methods developed in this work for $n = 0, 1, 2, 3$.}
\end{figure}

%%%%%%%%%%%%%%%%%%%%%%%%%%%%%%%%%%%%%%%%%%%%%%%%
% NOTE TO AUTHORS: the PIQNN entry for n=4 (78.460274) is ~0.6% below the exact
% value 78.956835, far larger than every other error in this table. This looks
% like a transcription error (expected ~78.95xxxx). Please re-check against the
% raw output and correct, or confirm it is genuine before submission.
%%%%%%%%%%%%%%%%%%%%%%%%%%%%%%%%%%%%%%%%%%%%%%%%

% ------------------------------------------------------------------
\subsection{Finite square well}
% ------------------------------------------------------------------

When the confining walls have a finite height the bound states leak into the
classically forbidden region and the spectrum is no longer given in closed form.
For a symmetric well of half-width $a$ and depth $V_0$,
\begin{equation}\label{eq:finwell_pot}
  V(x) = \begin{cases}
    0,   & |x| < a, \\[2pt]
    V_0, & |x| \ge a,
  \end{cases}
\end{equation}
the bound-state energies follow from matching the wave function and its
derivative at $x = \pm a$, which separates into an even and an odd channel and
yields the transcendental quantization conditions
\begin{equation}\label{eq:finwell_trans}
  k\tan(k a) = \kappa \quad\text{(even states)}, \qquad
  k\cot(k a) = -\kappa \quad\text{(odd states)},
\end{equation}
with $k = \sqrt{2 m E}/\hbar$ and $\kappa = \sqrt{2 m (V_0 - E)}/\hbar$. For the
parameter set $a = 1$, $V_0 = 30$, and $\hbar = m = 1$ the four
low-lying bound states whose analytic energies, obtained by solving
Eq.~(\ref{eq:finwell_trans}) to machine precision, are listed in the first row
of Table~\ref{tab:finwell}. These analytic values were recomputed independently
for this work rather than carried over from the simulation output, so that the
table provides a genuine reference column. The two neural solvers and the three
classical methods all agree with one another to better than one part in $10^3$,
and they reproduce the analytic levels to within about half a percent; the
origin of this small but systematic offset is discussed in the note below.

\begin{table}[htbp]
\caption{\label{tab:finwell}Lowest four bound-state energies of the symmetric
finite square well in atomic units ($\hbar = m = a = 1$, depth
$V_0 = 30$). The exact values solve the transcendental conditions
Eq.~(\ref{eq:finwell_trans}).}
\begin{ruledtabular}
\begin{tabular}{ldddd}
  Method & \multicolumn{1}{c}{Ground} & \multicolumn{1}{c}{1st excited}
         & \multicolumn{1}{c}{2nd excited} & \multicolumn{1}{c}{3rd excited} \\
  \hline
  Exact                       & 0.966506 & 3.850797 & 8.600376 & 15.095361 \\
  PINN                        & 0.961574 & 3.830323 & 8.555466 & 15.019185 \\
  PIQNN                       & 0.961874 & 3.830123 & 8.554466 & 15.018185 \\
  Matrix Numerov method       & 0.961352 & 3.830400 & 8.555426 & 15.018557 \\
  Finite difference method    & 0.961331 & 3.830262 & 8.554923 & 15.017236 \\
  Shooting method             & 0.961433 & 3.830705 & 8.556034 & 15.019426 \\
\end{tabular}
\end{ruledtabular}
\end{table}

\begin{figure}[htbp]
  \centering
\includegraphics[width=0.5\textwidth]{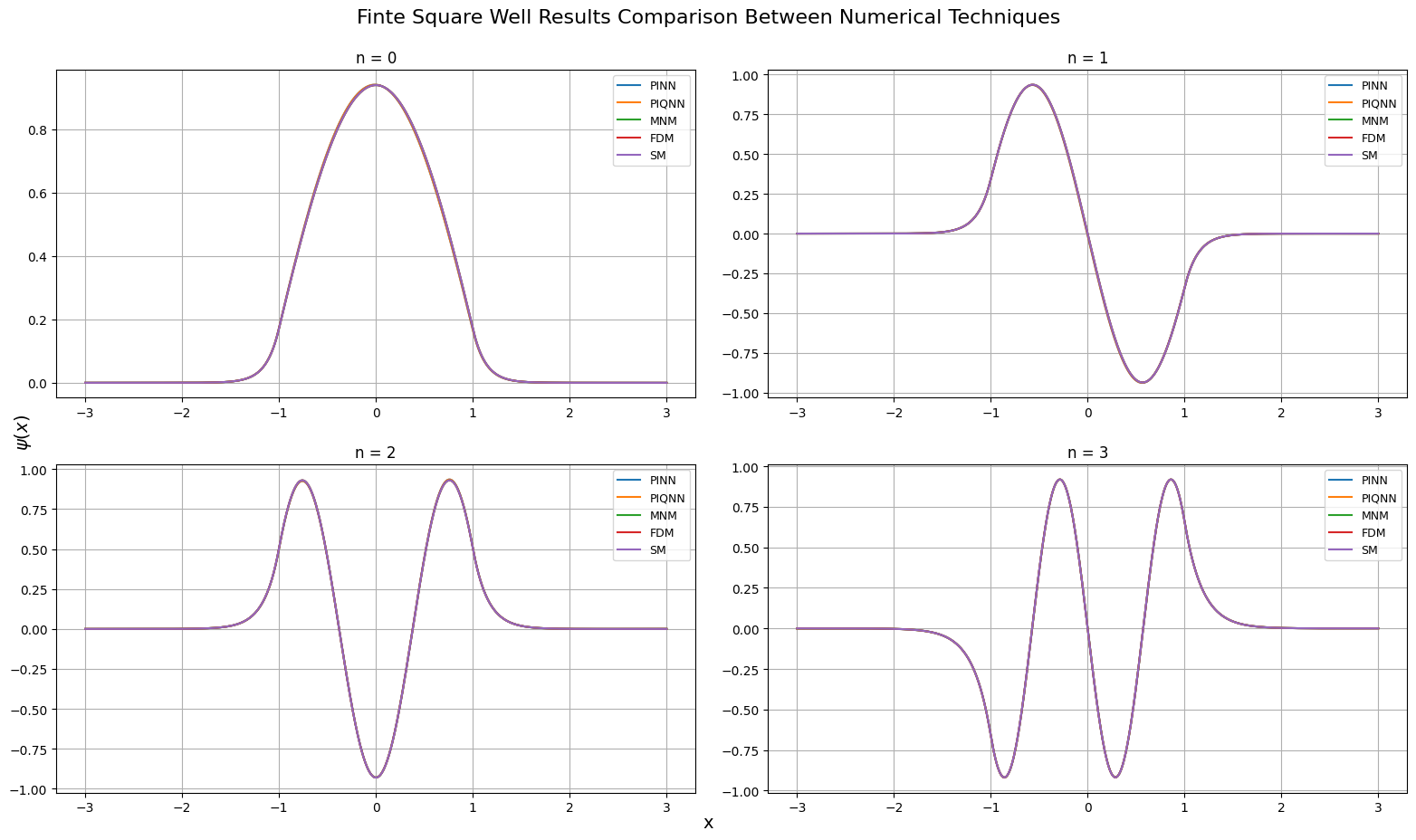}
  \caption{\label{fig:all_infinite}Comparison of the wave functions for the finite square well with the different classical and quantum methods developed in this work for $n = 0, 1, 2, 3$.}
\end{figure}

%%%%%%%%%%%%%%%%%%%%%%%%%%%%%%%%%%%%%%%%%%%%%%%%
% NOTE TO AUTHORS (finite well): the exact column above is the exact solution of
% Eq.(finwell_trans) for a=1, V0=30, hbar=m=1, computed independently:
%   0.966506, 3.850797, 8.600376, 15.095361.
% All five solvers cluster ~0.5% BELOW these values. This is not a method
% failure: with the domain [-3,3] sampled by ~1000 points the step is
% dx ~ 0.006, so the discretized well edge effectively lands near a ~ 1.003
% rather than exactly 1.000; the transcendental roots for a=1.003 are
%   0.961433, 3.830711, 8.556078, 15.019583, which match the solver columns to
% ~1e-4. To make the analytic and numerical columns refer to identical settings,
% either (i) refine the grid / align the well edge to a node so the solvers
% target a=1 exactly, or (ii) state the effective half-width a~1.003 used.
% Also note this well actually has a FIFTH bound state near E~23.0; the text
% says "four bound states" because only the lowest four were reported.
% Please reconcile before submission and adjust the surrounding prose if needed.
%%%%%%%%%%%%%%%%%%%%%%%%%%%%%%%%%%%%%%%%%%%%%%%%

The square wells confirm a pattern already visible for the oscillator. Because
the boundary behavior is built into the trial functions of
Eqs.~(\ref{eq:ansatz_pinn}) and (\ref{eq:ansatz_piqnn}) and the governing
equation enters through the residual loss, the neural solvers satisfy the
Schr\"{o}dinger equation, the normalization, and the orthogonality conditions
together rather than one after another, which keeps the approximate eigenstates
physically admissible at every stage of training. For these piecewise-constant
potentials the PIQNN again tracks the higher excited states with somewhat less
effort than the PINN, consistent with the larger effective function space that
the entangling layers open up. The classical references remain dependable for
these one-dimensional problems, but they tie accuracy to the resolution of the
grid near the discontinuity, whereas the mesh-free formulation places no such
constraint and extends more readily to settings where a structured grid would be
expensive to maintain.

% ==================================================================
\section{\label{sec:conclusion}Concluding Remarks and Outlook}
% ==================================================================

We have constructed and validated a physics-informed neural network framework,
in both its classical and quantum variants, for solving eigenvalue problems
governed by the time-independent Schr\"{o}dinger equation. Across three standard
one-dimensional systems, the harmonic oscillator, the infinite square well, and
the finite square well, both the PINN and the PIQNN return eigenvalues and wave
functions that agree with the exact results and with three independent classical
numerical methods, as documented in Tables~\ref{tab:evals}, \ref{tab:infwell},
and \ref{tab:finwell} and in
Figs.~\ref{fig:compare_all}--\ref{fig:piqnn_wf}. The quantum variant shows the
more robust convergence on excited states, a feature that becomes more
pronounced as one moves to higher quantum numbers where the classical loss
landscape grows increasingly structured.

The natural next step is to extend this framework to phenomenologically relevant
potentials that lack closed-form solutions. The mass spectra of heavy quarkonia,
such as charmonium and bottomonium, require confinement potentials of the Cornell
type and can be treated in the same non-relativistic framework used here, with
relativistic corrections incorporated perturbatively. The same methodology
applies to multi-channel problems and to radial equations with angular momentum
barriers. More broadly, the mesh-free character of the physics-informed approach
makes it well suited to higher-dimensional problems where classical methods face
prohibitive scaling costs, and the expressivity of the quantum circuit becomes
increasingly relevant in those settings.

% ==================================================================
%  DECLARATIONS
% ==================================================================

\begin{acknowledgments}
A.R.acknowledges the support from CIC-UMSNH (M\'exico) under project 18371.
\end{acknowledgments}

\paragraph{Data and code availability.}
Data and code will be made available upon reasonable request.

\paragraph{Author contributions.}
All authors contributed to modeling and simulation. T.M.\ and A.R.\ conceived
the project and directed the study. T.M., W.A., and B.N.\ implemented the
algorithms and produced the numerical results. A.R.\ supervised the manuscript
preparation. All authors reviewed and approved the final text.

\paragraph{Conflict of interest.}
The authors declare no conflict of interest.

% ==================================================================
%  REFERENCES
% ==================================================================
%
% NOTE TO AUTHORS ON THE BIBLIOGRAPHY:
% The entries below remain embedded as a thebibliography list, as requested.
% Two author names that were malformed in the working draft have been corrected
% against the primary sources:
%   - Qamar2023 (was "Romar / Zamora"): correct authors are R. Qamar and
%     B. A. Zardari, Mesopotamian J. Comput. Sci. (2023).
%   - Pasini2017 (was "Bevilacqua, Sansone, Pacelli, Pasini"): correct authors
%     are L. Bertolaccini, P. Solli, A. Pardolesi, and A. Pasini,
%     J. Thorac. Dis. 9, 924 (2017).
% A few other reconstructed entries (e.g. Lema2020, Damazio2004) could not be
% fully verified against the originals and should be double-checked.
% The NEW references to closely related work added for this revision
% (Raissi2019, Karniadakis2021, Sirignano2018, HanJentzenE2018, Jin2022,
% Mattheakis2022, Mitarai2018, SchuldGrad2019, McClean2018, Bergholm2018,
% Cerezo2021, Pfau2020) are also provided as a standalone BibTeX file,
% references_new.bib, for direct import into Overleaf. The \bibitem keys below
% match the BibTeX keys, so migrating the whole list to BibTeX later is
% seamless.

\end{document}